\documentclass[twocolumn,aps,pra]{revtex4-1}

\usepackage{forloop,ifthen} 
\usepackage{multirow} 
\usepackage{amsmath} 
\usepackage{ esint } 
\usepackage{graphicx} \usepackage{subfigure} \usepackage{color}
\usepackage{dcolumn}
    \usepackage{bm}
    \usepackage{xspace}
    \usepackage{hyperref}

\newcommand{\WF}{Wigner current\xspace}
\newcommand{\SP}{stagnation point\xspace}
\newcommand{\CO}[1]{}{}
\newcommand{\emphCaption}[1]{{{#1}}}
\newcommand{\VH}{V^{\odot}}

\newcommand{\VE}{V^{\cal E}}
\newcommand{\VM}{V^{\cal M}}
\newcommand{\VRM}{V^{\cal R}}
\newcommand{\VV}{V^{\cal A}}

\begin{document}

\title{Wigner's quantum phase-space current in weakly-anharmonic\\  weakly-excited two-state systems}

\author{Dimitris Kakofengitis and Ole Steuernagel}

\affiliation{School of Physics, Astronomy and Mathematics, University of Hertfordshire, Hatfield, AL10 9AB, UK }

\date{\today}

\begin{abstract}
  {There are no phase-space trajectories for anharmonic quantum systems, but Wigner's
    phase-space representation of quantum mechanics features Wigner current~$\bm J$. This
    current reveals fine details of quantum dynamics -- finer than is ordinarily thought
    accessible according to quantum folklore invoking Heisenberg's uncertainty
    principle. Here, we focus on the simplest, most intuitive, and analytically accessible
    aspects of~$\bm J$. We investigate features of~$\bm J$ for bound states of
    time-reversible, weakly-anharmonic one-dimensional quantum-mechanical systems which
    are weakly-excited. We establish that weakly-anharmonic potentials can be grouped into
    three distinct classes: hard, soft, and odd potentials.  We stress connections between
    each other and the harmonic case. We show that their Wigner current fieldline patterns
    can be characterised by~$\bm J$'s discrete stagnation points, how these arise and how
    a quantum system's dynamics is constrained by the stagnation points' topological
    charge conservation. We additionally show that quantum dynamics in phase space, in the
    case of vanishing Planck constant $\hbar$ or vanishing anharmonicity, does not
    pointwise converge to classical dynamics.}
\end{abstract}



\maketitle

\section{Introduction \label{sec_intro}}

Classical phase-space trajectories allow the viewer to characterise a system's dynamics
at a glimpse. They also reveal rich structures such as the strange attractors of
chaotic systems full of intricacies and
beauty~\cite{Berry_AIPC78,Cvitanovic_Chaos_book_12}.

Anharmonic quantum systems do not feature
trajectories~\cite{Oliva_Traj1611}, but fieldlines of Wigner's
phase-space current~$\bm J$ characterise quantum-mechanical phase-space dynamics at a
glimpse (Sections~\ref{sec_5_WFpatterns_EVs} and~\ref{sec_6_2state_dynamics}), similar
to classical phase portraits: this is underexplored.

This is a gap this work aims to help fill.

Wigner's quantum theory~\cite{Wigner_PR32} (Section~\ref{sec_2_WF}) is a representation of
quantum mechanics in phase
space~\cite{Zachos_IJMP02,Hirshfeld_AJP02,Hancock_EJP04,Rasinariu_FP12,Zachos_book_05,Hillery_PR84,Case_AJP08,Kakofengitis_PRA17}
(additionally pioneered by Groenewold~\cite{Groenewold_Phys46} and
Moyal~\cite{Moyal_MPCPS49}) equivalent to Heisenberg, Schr\"odinger and Feynman's
representations of quantum theory. It is, historically speaking, the third representation
of quantum physics and its importance is still not clear: ``\emph{Some believe it will
  supplant, or at least complement, the other methods in quantum mechanics and quantum
  field theory}"\cite{Hirshfeld_AJP02}.

Here we investigate Wigner's quantum phase-space current~$\bm J$ and its fieldlines for
the three classes of weakly-anharmonic potentials: hard, soft and odd potentials
(Section~\ref{sec_4_Anharm_pot}). We emphasize the features the current patterns,
associated with the three classes of potentials, have in common: it turns out that odd
potentials are hybrids of hard and soft potentials and this is reflected in their
phase-space current patterns~(Sections~\ref{sec_5_WFpatterns_EVs}
and~\ref{sec_6_2state_dynamics}).

We particularly stress an intuitive understanding of how the Wigner current patterns
emerge~(Subsections~\ref{subsec_WF_points}
and~\ref{subsec:QualitativeAnharmonic}). Because the anharmonicities are weak, the $\bm
J$-fieldline patterns can partly be understood from the vantage point of the harmonic
oscillator~(Section~\ref{sec_3_Harm_case}) and partly through perturbation analyses
(Sections~\ref{paragraph_Displ_min_vortex} and~\ref{paragraph_Displ_Jp_Jx}).

For simplicity, our discussions are limited to one-dimensional conservative
quantum-mechanical systems featuring nearly harmonic potentials. We only consider the
bound energy eigenstates~(Section~\ref{sec_5_WFpatterns_EVs}) of weakly-excited systems in
pure two-state superpositions~(Section~\ref{sec_6_2state_dynamics}).

To demonstrate the conceptual power of the use of~$\bm J$ and collections of its
fieldlines, we show that in the limit of vanishing anharmonicity the fieldlines of~$\bm J$
do not converge pointwise~(Section~\ref{sec_5_WFpatterns_EVs}
and~\ref{sec_6_2state_dynamics}) to those of the harmonic
oscillator~(Section~\ref{sec_3_Harm_case}). This implies that in the limit of vanishing
anharmonicity, or vanishing magnitude of Planck's constant, quantum and classical
phase-space behaviour are qualitatively very different from each
other~\cite{Oliva_Traj1611,Kakofengitis_PRA17}, see
Sections~\ref{subsec:QualitativeAnharmonic} and~\ref{sec_7_conclusions}.

\section{Wigner distributions and Wigner current \label{sec_2_WF}}

We parameterize pure quantum two-state superpositions of energy eigenstates $\psi_m$ (with
eigenenergies~$E_m$ and energy difference~$\Delta E=E_n-E_m$) by the mixing angle~$\theta$
\begin{equation}\label{eq:superposition_state}
  \Psi_{m,n}(x,t;\theta)=\cos(\theta)\psi_m(x)+\sin(\theta)e^{-\frac{i}{\hbar}\Delta E t} \psi_n(x)\; .
\end{equation}
Here, $x$ and~$t$ denote position and time, $\hbar=h/(2\pi)$ is Planck's constant, and
since this is a two-state superposition, the period time is
\begin{equation}\label{eq:RespPeriod}
  T_{m,n}=\frac{2\pi\hbar}{\Delta E}=\frac{2\pi\hbar}{|E_n-E_m|}\; .
\end{equation}
Wigner's phase-space quantum distribution
$W(x,p,t)$~\cite{Wigner_PR32,Hillery_PR84,Case_AJP08} is
\begin{equation}\label{eq:W}
  W(x,p,t) \equiv \frac{1}{\pi \hbar} \int
  dy \, \varrho(x-y,x+y,t)e^{\frac{2i}{\hbar} p y}\; ;
\end{equation}
where~$ \varrho(x-y,x+y,t) = \Psi(x-y)\Psi^*(x+y)$, for pure states.
$W$ is real-valued, non-local (through~$y$), and
\begin{figure*}[t]
  \centering
  \includegraphics[width=0.8\textwidth]{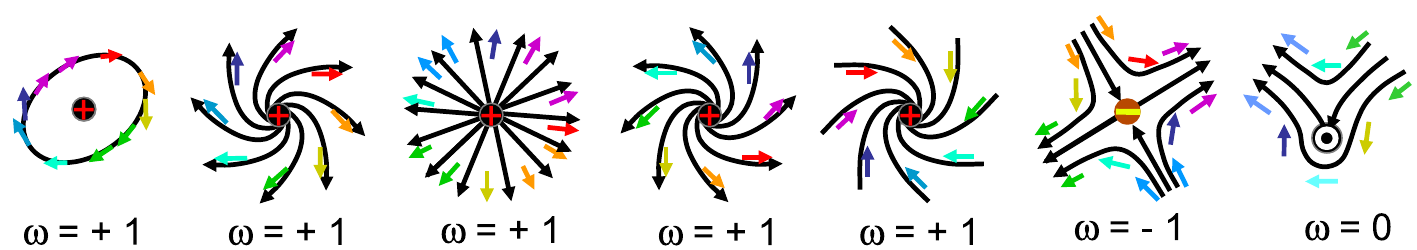}
  \caption{\CO{(Color Online)} \emphCaption{Wigner current stagnation points are
      characterised} by their integer orientation winding
    number~(\ref{eq:WindingNumber}).
 A red plus sign labels stagnation points with topological 
charge~\cite{Dennis_PO09}~$\omega=+1$, a yellow minus sign~$\omega=-1$, 
and a  white circle~$\omega=0$. The current's fieldlines can be skewed near 
stagnation points in phase-space, can feature skewed separatrices, and saddles
    oriented in the $p$-direction. 
    \label{fig:stagnation_points_omega}}
\end{figure*}
normalized $ \int \int dx \; dp \; W(x,p,t) = 1 $ (we abbreviate $\int = \int_{-\infty}^{\infty}$).

Wigner's distribution~$W$ is set apart from other quantum phase-space
distributions~\cite{Hillery_PR84} by the fact that only~$W$ simultaneously yields the
correct projections in position and momentum ($\varrho(x,x,t)= \int
dp \; W$ and $\tilde \varrho(p,p,t)=\int
dx \; W$) as well as state overlaps $|\langle \psi_1 | \psi_2 \rangle |^2 = 2 \pi \hbar
\int
\int
dx\; dp \; W_1 \; W_2$, while maintaining its form~(\ref{eq:W}) when evolved in
time. Additionally, the Wigner distribution's averages and uncertainties evolve
\emph{momentarily} classically~\cite{Royer_FOP92,Ballentine_PRA94} (fulfilling Ehrenfest's
theo\-rem~\cite{Kakofengitis_PRA17}). This is why~$W$ is the \emph{``closest quantum
  analogue of the classical phase-space distribution''}~\cite{Zurek_NAT01}.

To study~$W$'s dynamics one Wigner-transforms von~Neumann equation $ \partial_t
\varrho = - \frac{i}{\hbar} [H,\varrho]$, analogously to Eq.~(\ref{eq:W}).  The result can
be cast into the form of Wigner's continuity equation~\cite{Wigner_PR32}
\begin{equation} \partial_t W + \partial_x J_x + \partial_p J_p = 0 \; ,
\label{eq:W_Continuity}
\end{equation} 
here $\bm J(x,p,t)$ denotes the Wigner current~\cite{Ole_PRL13} and the shortened notation
$\frac{\partial^2}{\partial x^2}=\partial_x^2$, etc., is used for partial derivatives.

In the case of potentials that can be Taylor-expanded, $\bm J$ assumes the infinite-sum
form~\cite{Wigner_PR32,Groenewold_Phys46,Moyal_MPCPS49,Donoso_PRL01,Schleich_01,Case_AJP08,Bauke_2011arXiv1101.2683B}
\begin{multline}
  \label{eq:CurrentComponents}
  \bm{J}=\begin{pmatrix}J_x\\J_p\end{pmatrix}= \begin{pmatrix}\frac{p}{M}W\\
    -\sum\limits_{l=0}^{\infty}{\frac{(i\hbar/2)^{2l}}{(2l+1)!}
      \partial_p^{2l} W \partial_x^{2l+1} V(x) }\end{pmatrix}\; ,
\end{multline}
where $M$ is the mass of the particle.  The term~$J_{p,l=0}$ is of classical form, the
terms of higher order in $l$ are known as quantum correction terms.

In general~$J_p$ in~(\ref{eq:CurrentComponents}) has the integral
form~\cite{Wigner_PR32,Kakofengitis_PRA17}
\begin{multline}
  \label{eq:Wigner_Current_Jp_Integral_Form}
  J_p=-\frac{1}{\pi\hbar}\int dy\; \left[\frac{V(x+y)-V(x-y)}{2y}\right]\\
  \times\varrho(x-y,x+y,t) e^{\frac{2i}{\hbar}py}\; .
\end{multline}
For numerical stability we avoid the infinite-sum form of $J_p$
in~(\ref{eq:CurrentComponents}), when generating the $\bm J$'s fieldlines depicted in
Figs.~\ref{fig:zeros_anharm_1},~\ref{fig:EckartImpPlots},~\ref{fig:RosenMorse_ZoomIn}
and~\ref{fig:zeros_2nd_Excited_state}, and use its integral
form~(\ref{eq:Wigner_Current_Jp_Integral_Form}) instead.

\subsection{Stagnation points of quantum phase-space
  current \label{subsec:J_StagnationPoints}}

Wigner current reveals detail of quantum dynamics' finest features, in particular the
nature of its \emph{stagnation points}, the points in phase-space where the dynamics
momentarily stops. In classical physics, stagnation points are also referred to as
equilibrium, stationary, fixed, critical, invariant and rest
points~\cite{Cvitanovic_Chaos_book_12}. This multitude of terms testifies to their central
importance in classical mechanics as well as wave theory, e.~g. in the field of `singular'
optics~\cite{Dennis_PO09} (where they are called singular points~\cite{Berry_NAT00}).

In classical physics stagnation points can only form on the $x$-axis at the
potential's minima, maxima and saddle points, where the force is zero.

In quantum physics Wigner distributions are known to feature negative regions in
phase-space~\cite{Wigner_PR32}. In these regions the current is inverted in
direction~\cite{Ole_PRL13}, this leads to the formation of whorls and saddle flows with
\emph{points} of stagnating current at their centers. This can happen wherever in
phase-space the Wigner distribution turns negative.

Analogously to the classical case, \emph{Wigner current's stagnation points} are the most
important points in quantum phase-space for two reasons: the topological nature of \WF's
{\SP}s, firstly, orders the current in large surrounding sectors of phase-space and,
secondly, makes them carry a conserved topological charge. Its topological nature makes
their appearance robust to perturbations and time evolution.

To quantify the topological charge conservation of~$\bm J$'s stagnation points, we use the
orientation winding number for Wigner current along closed, self-avoiding loops~$\cal L$
in phase-space~\cite{Ole_PRL13}
\begin{eqnarray}
  \label{eq:WindingNumber}
  \omega({\cal L},t) =\frac{1}{2\pi} \varointctrclockwise_{\cal L} d
  \varphi \; ;
\end{eqnarray}
where~$\varphi$ is the angle between~$\bm J$ and the~$x$-axis, see
fig.~\ref{fig:stagnation_points_omega}. Since the components of the current are continuous
functions, $\omega$ is zero except for the case when the loop~$\cal L$ contains stagnation points,
such as those sketched in fig.~\ref{fig:stagnation_points_omega}. Then, a non-zero value
of $\omega$ can occur. The value of~$\omega$, (in mathematics known as the stagnation
point's Poincar\'e-Hopf \emph{index}), is topologically protected~\cite{Dennis_PO09}. When
the system's dynamics transports a stagnation point across~$\cal L$, $\omega({\cal L},t)$
can change~\cite{Ole_PRL13}. The topological charges can be combined or split through the
system's time evolution while their sum remains conserved~\cite{Ole_PRL13}.

\section{Wigner current of Harmonic Oscillators \label{sec_3_Harm_case}}
\begin{figure}[b]
  \centering
  \includegraphics[width=0.45\textwidth]{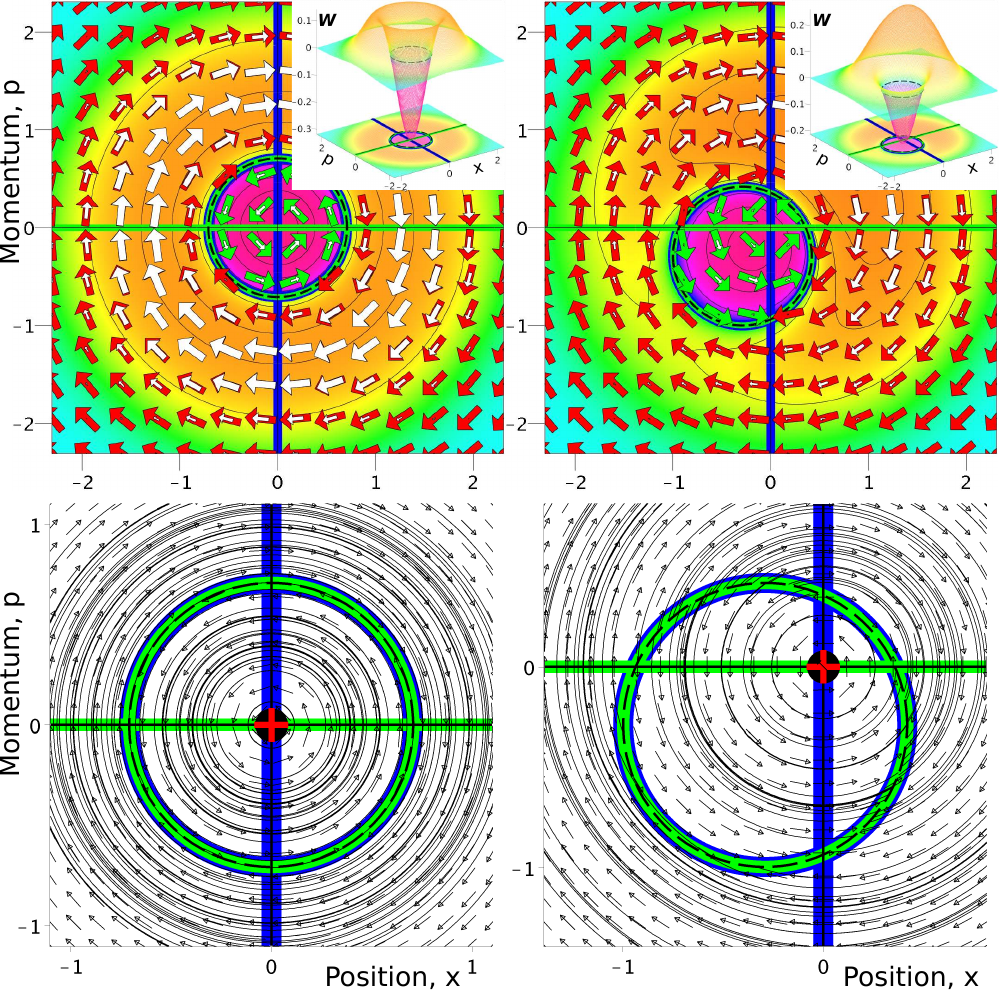}
  \caption{\CO{(Color online)} \emphCaption{Top row: Wigner current and Wigner distributions for the
      harmonic oscillator potential~(\ref{eq:Vharm}). Bottom row: Fieldlines of the integrated
      Wigner current~$\bm J(x,p,t)$ at fixed time~$t$}. Left column: first excited
    state~$\psi_1$. Right column: the superposition
    $\Psi^\odot_{0,1}(\frac{7T}{8};\frac{\pi}{3})$. Dashed black circles show the locations of $W=0$
    which implies that, because of Eq.~(\ref{eq:WF_harm}), $J^\odot_x=0$~(thick green lines) and
    $J^\odot_p=0$~(thick blue lines). These dashed black lines constitute circular \emph{current
      stagnation lines} with constant radius $R^\odot_{ 1}=1/\sqrt{2}$ and a centre which is further
    displaced from the origin~(\ref{eq:HarmCentr}) the smaller the mixing angle $\theta$
    in~(\ref{eq:superposition_state}). In the top row the background coloring refers to the
    respective Wigner distributions' values (compare insets). The normalised Wigner current $\bm
    J/|\bm J|$ is depicted with red arrows if the Wigner distribution is positive. For negative
    Wigner distribution green arrows are used, demonstrating \emph{current
      reversal}~\cite{Ole_PRL13}. White arrows on top of~$\bm J/|\bm J|$ depict the magnitude
    of~$\bm J$.
    Bottom row: The depicted Wigner current fieldlines run through randomly picked points. The
    stagnation point at the origin carries topological charge~$\omega=+1$ and is labelled as in
    fig.~\ref{fig:stagnation_points_omega}. \label{fig:current_harm}}
\end{figure}

When we introduce weakly-anharmonic potentials in section~\ref{sec_4_Anharm_pot}, we
rescale them to match their minimum's curvature to our choice of a harmonic reference
potential
\begin{equation}
  \VH(x) = \frac{x^2}{2} \; ,
  \label{eq:Vharm}
\end{equation}
with \emph{circular} fieldlines~(rather than elliptical~\cite{Dahl_JCP88}), see
fig.~\ref{fig:current_harm}. Having such circular fieldlines is the main motivation for
this particular choice. It constitutes a choice of units of mass~$M=1$, spring
constant~$k=1$. Setting $\hbar=1$, leads to an angular frequency of $\Omega^\odot=1$
and an oscillation period of $ T^\odot = {2\pi}$.

Wigner current for~$\VH$, according to Eq.~(\ref{eq:CurrentComponents}), has the
`classical' form (see Takabayasi~\cite{Takabayasi_PTP54}~p.351)
\begin{equation} {\bm J}^\odot = W^\odot(x,p,t) \cdot \binom{ p }{ -x } \, .
  \label{eq:WF_harm}
\end{equation}

\subsection{Degenerate Wigner current \label{subsec:hosc_J_degenerate}}

Wigner distributions are continuous and have negativities~\cite{Wigner_PR32}, they therefore feature
zero-contours in phase-space which, in the case of the harmonic oscillator, because of the
form~(\ref{eq:WF_harm}) of $\bm J$, become zero-lines for both components,~$J^\odot_x$ and
$J^\odot_p$ simultaneously, giving rise to \emph{lines of stagnation} of the current, see
fig.~\ref{fig:current_harm}.

It is well known from quantum optics that the eigenstates~$W^\odot_{n,n}(x,p)$ of the
harmonic oscillator Fock states~$\psi_n^\odot$ resemble Mexican hats centred on the
origin, with concentric fringes of alternating polarity. Their zero-contours thus form
concentric circles~\cite{Schleich_01}.

In the case of superposition states, we primarily investigate superpositions of ground and
first excited state $\Psi_{0,1}(t;\theta)$~(\ref{eq:superposition_state}), in which case
$W$'s circular zero-contour remains a circle with constant radius $R^\odot_{
  1}=1/\sqrt{2}$ but shifted centre position~$C_{01}$. The centre is further displaced
from the origin the larger the groundstate contribution ($\theta \downarrow 0$ in
$\Psi_{0,1}(\theta)$~(\ref{eq:superposition_state})) and rotates around the origin, with
frequency~$\Omega^\odot = 1$, according to
\begin{equation}\label{eq:HarmCentr}
  C_{0,1}(t;\theta)= \left(-R^\odot_{1}\cos(t)\cot(\theta) ,
    R^\odot_{1}\sin(t)\cot(\theta)\right)\; ,
\end{equation}
compare Figs.~\ref{fig:current_harm} and~\ref{fig:SeveralRabiCycles}.

\section{The Three Classes of Weakly-Anharmonic Potentials
\label{sec_4_Anharm_pot}}

Weakly-anharmonic potentials $V(x)$ that admit a Taylor expansion in~$x$ are characterized
by their leading anharmonic term~$\alpha_\nu x^\nu$ in what we will refer to as their
\emph{truncation}~$\VV_{\nu}$ of order $\nu$, and representative~$\cal A$, namely,
\begin{equation}
  V^{\cal A}(x) \approx  \VV_{\nu}(x) =  \frac{x^2}{2} 
   + \alpha_\nu^{\cal A} x^\nu  =  \VH(x) + \alpha_\nu^{\cal A} x^\nu \, . 
  \label{eq:V_trunc}
\end{equation}

\begin{table*}[t!]
  \centering
  \begin{tabular}{l|cccc}
    Potential & 
    Harmonic Oscillator&Eckart (hard)&Rosen-Morse (soft)&Morse (odd)\\[2mm]
    \hline\\[-3mm]
    $V^{\cal A}(x)$& $V^\odot=\frac{x^2}{2}$& $V^{\cal
      E}=D\tan^2\left(\frac{x}{\sqrt{2D}}\right)$& $V^{\cal
      R}=D\tanh^2\left(\frac{x}{\sqrt{2D}}\right)$& $V^{\cal
      M}=D\left(1-e^{-\frac{x}{\sqrt{2D}}}\right)^2$
    \\[2mm]
    $V_\nu^{\cal A}=\frac{x^2}{2}+\alpha_\nu^{\cal A} x^\nu$& $V^\odot=\frac{x^2}{2}$&
    $V_4^{\cal E}=\frac{x^2}{2}+\frac{x^4}{6D}$& $ V_4^{\cal R}=\frac{x^2}{2}-\frac{x^4}{6D}$&
    $ V_3^{\cal M}=\frac{x^2}{2}-\frac{x^3}{2\sqrt{2D}}$
    \\[1mm]
    Eigenvalues& $n+\frac{1}{2}$&
    $D\left[\left(\frac{\sqrt{1+16D^2}+2n+1}{4D}\right)^2-1\right]$&
    $D\left[1-\left(\frac{\sqrt{1+16D^2}-2n-1}{4D}\right)^2\right]$&
    $D\left[1-\left(\frac{4D-2n-1}{4D}\right)^2\right]$
    \\[1mm]
    $\mbox{Groundstate\quad}\atop\mbox{(unnormalized)}$& $e^{-\frac{x^2}{2}}$&
    $\cos\left(\frac{x}{\sqrt{2D}}\right)^{\left[\frac{\sqrt{1+16D^2}+1}{2}\right]}$& ${\rm
      sech}\left(\frac{x}{\sqrt{2D}}\right)^{\left[\frac{\sqrt{1+16D^2}-1}{2}\right]}$&
    $e^{-\left[x\left(\frac{4D-1}{2\sqrt{2D}}\right)+2De^{-\frac{x}{\sqrt{2D}}}\right]}$
    \\[2mm]
    $D$-parameter& -& $D>0$& $D>\sqrt{\left(\frac{2N-1}{4}\right)^2-\frac{1}{16}}$&
    $D>\frac{2N-1}{4}$
    \\[2mm]
    $N$ bound states& -& -& $N(D)=\left\lfloor\frac{\sqrt{1+16D^2}+1}{2}\right\rfloor$&
    $N(D)=\left\lfloor\frac{4D+1}{2}\right\rfloor$
    \end{tabular}
    \caption{The quantum harmonic oscillator and the three classes of anharmonic
      potentials are presented together with their corresponding energy eigenvalues and
      groundstate~\cite{Dutt_AJP88} ($n=0,1,2,\ldots,N-1$). The inverse anharmonicity
      parameter~$D$ ($\lim_{D \rightarrow \infty}V^{\cal E,R,M}=\VH$) is also the depth of the
      two open potentials~($V^{\cal R}$ and $V^{\cal M}$).
      The last row gives the number of bound states~$N$ in terms
      of~$D$, where~$\left\lfloor\ldots\right\rfloor$ denotes floor rounding. Higher
      excited energy eigenstates are easily created from the groundstates using SUSY-QM
      techniques~\cite{Dutt_AJP88}.
      \label{tab:SIPs}}
\end{table*}

The precise order~$\nu$ of a truncation's leading anharmonic term~$\alpha_\nu^{\cal A}
x^\nu$ is quite unimportant for our discussion, as it is the qualitative class of the
potential that determines its qualitative dynamic features we are primarily interested in.

\begin{figure}[t]
  \centering
  \includegraphics[width=0.35\textwidth]{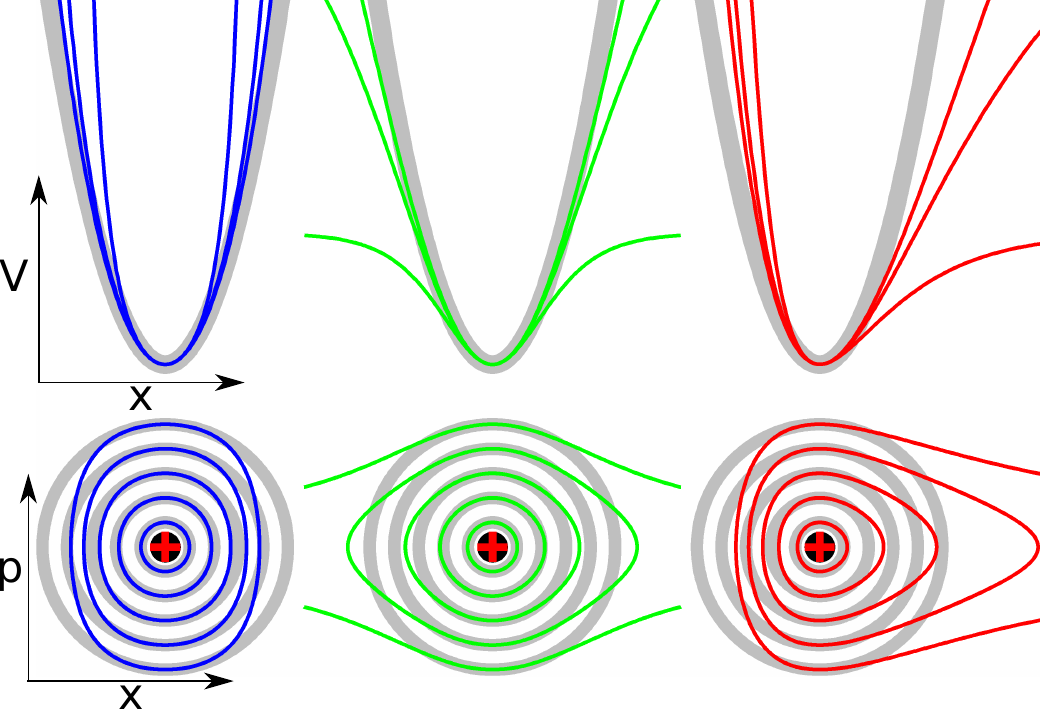}
  \caption{\CO{(Color Online)} {Top row:} Representatives of the three classes of weakly
    anharmonic potentials displayed side by side: Eckart potentials ($\VE$, left, in blue)
    are hard, Rosen-Morse potentials ($\VRM$, centre, in green) are soft and Morse
    potentials ($\VM$, right, in red) are odd, see~Table~\ref{tab:SIPs}.  The potentials
    feature differing amounts of anharmonicity,~$\alpha_\nu^{\cal A}$~(\ref{eq:V_trunc}),
    while all are rescaled to have the same minimum curvature as the harmonic reference
    potential $\VH$~(\ref{eq:Vharm}) (displayed behind each class as a thick grey line).
    \\
    {Bottom row:} energy contours (for one fixed potential strength in each column)
    superimposed on harmonic oscillator's circular energy contours (thick grey
    lines).\label{fig:Potential}}
\end{figure}

With respect to qualitative features of Wigner current for weakly-excited bound state
systems, just as for the associated phase portraits in the classical case,
see~fig.~\ref{fig:Potential}, only three classes of anharmonic potentials exist:
\emph{hard, soft}, and \emph{odd} potentials. We checked this numerically for several
potentials and it is plausible from our discussion below.

All potentials with a leading positive anharmonic term of even order have qualitatively
similar classical phase-space profiles. They correspond to springs harder than their
Hookian reference~(\ref{eq:Vharm}), see left column of fig.~\ref{fig:Potential}.  Soft
potentials have a negative leading term of even order, fig.~\ref{fig:Potential}, middle
column. For potentials of a leading term of odd order we always set the leading term
$\alpha_\nu < 0$, making the odd potentials soft for $x>0$ and hard for $x<0$,
fig.~\ref{fig:Potential}, right column.

For each class a representative exists for which all bound state eigenfunctions and
eigenenergies are known in simple closed form~\cite{Dutt_AJP88}. As such representatives
we choose the hard Eckart,~$V^{\cal E}$, soft Rosen-Morse,~$V^{\cal R}$, and odd Morse
potential,~$V^{\cal M}$; see fig.~\ref{fig:Potential} and Table~\ref{tab:SIPs}. For the
Morse case all bound-state Wigner distributions are known~\cite{Dahl_JCP88} and were used
to cross-check some of our numerical calculations.

Anharmonic potentials $\VV$ which, based on their truncation $\VV_{\nu}$, are classed as
even or odd can contain higher order Taylor terms which are not necessarily only even or
odd. The influence of such higher terms can be neglected since we limit our investigation
to weakly excited systems. If we were to regard the truncated right hand side of
Eq.~(\ref{eq:V_trunc}) as the full potential, soft and odd potentials would obviously have
no bound eigenstates; we exclude such cases.

With these provisions, studying one representative of each class allows us to cover
qualitative features of Wigner current of the bound states of \emph{all} weakly-excited
weakly-anharmonic potentials.

\section{Wigner current patterns for Eigenstates of anharmonic
  potentials\label{sec_5_WFpatterns_EVs}}

\begin{figure}[b]
  \centering
  \includegraphics[width=0.26\textwidth,height=0.285\textwidth]{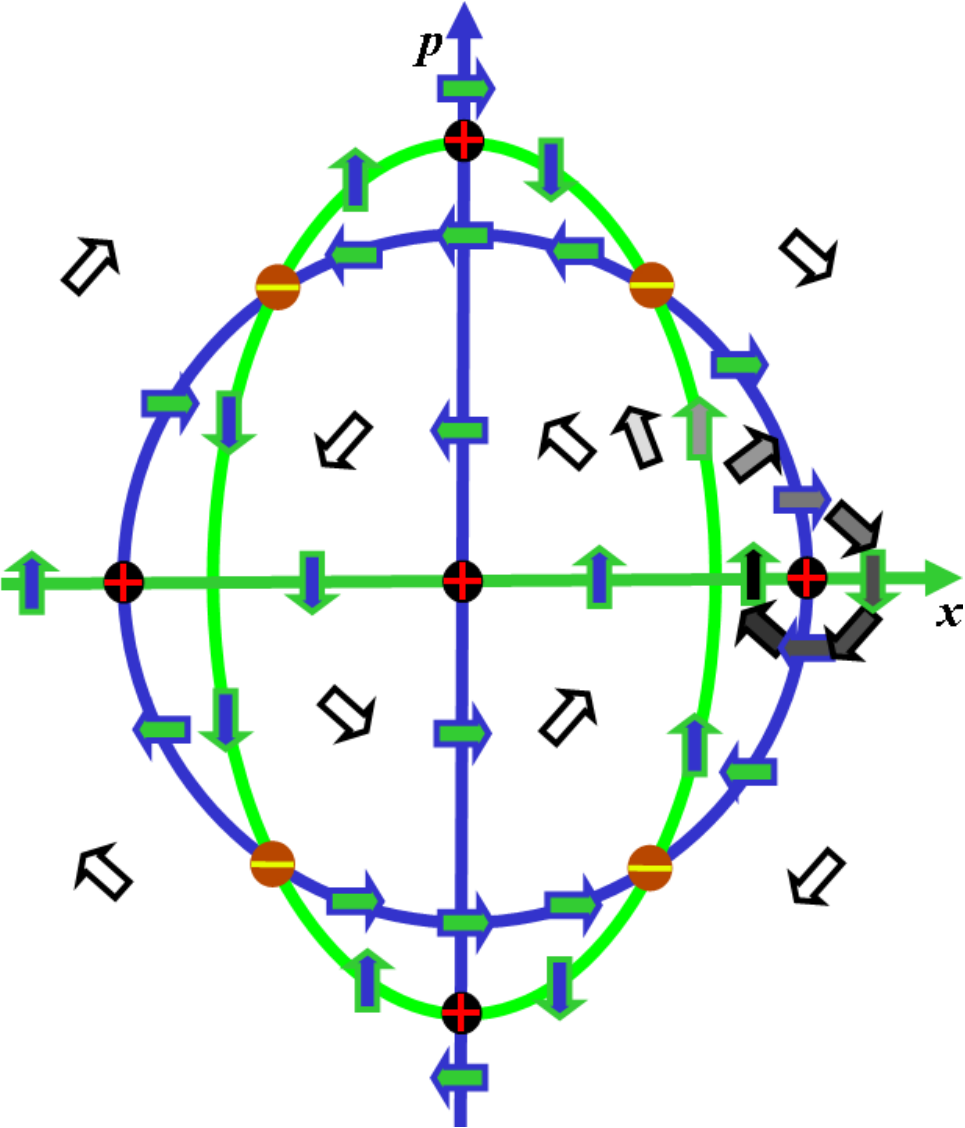}
  \caption{\CO{(Color Online)} \emphCaption{Qualitative discussion of emergence of type
      and positioning of Wigner current stagnation points for first excited state of a hard
      potential.}  White arrows represent normalized Wigner current as observed in the
    harmonic case; clockwise, far from the origin and, anti-clockwise (inverted current),
    very close to the origin (compare left panels of fig.~\ref{fig:current_harm}). Green and
    blue ellipses represent the zero-circles, see Section~\ref{subsec:hosc_J_degenerate},
    deformed by the anharmonicity of the potential. They are colored green when $J_x$
    changes sign and blue when $J_p$ does, this also applies to $x$ and $p$-axis.
    To `fill the plane' we track the orientation of the current while moving
    across phase-space along the sequence of arrows with ever darker shades
    of grey which eventually wraps around the rightmost `+1'-stagnation point.  Whenever
    this sequence crosses a zero-line, when $J_x=0$ or $J_p=0$, the arrows are framed
    green or blue, respectively. We can, similarly, track the current's orientation around the boundaries of the
    deformed zero-circles and along $x$ and $p$-axis. Green arrows with blue fringe are
    orientated horizontally ($J_p=0$) and invert direction whenever the blue line they are
    pinned to crosses a green line. Blue arrows with green fringe are tied to green lines,
    are vertically aligned, and behave analogously.
    At every crossing of a green with a blue line a stagnation point exists, but nowhere
    else. Having `filled the plane' we can work out the topological charge of the
    stagnation points, labelled as the symbols of
    fig.~\ref{fig:stagnation_points_omega}. The quantitative plots in the top row of
    fig.~\ref{fig:zeros_anharm_1} confirm this qualitative
    analysis.\label{fig:prediction_zeros_anharm}}
\end{figure}

\begin{figure*}[t]
 \centering
 \setlength{\fboxrule}{1pt}\setlength{\fboxsep}{1pt}
  \fbox{\includegraphics[width=0.485\textwidth]{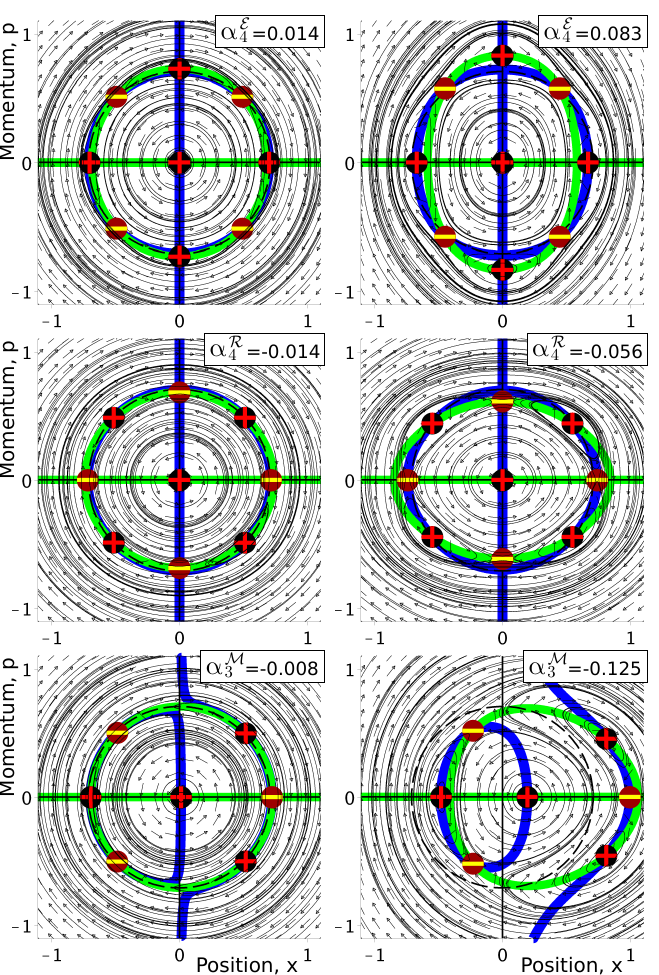}} 
  \fbox{\includegraphics[width=0.485\textwidth]{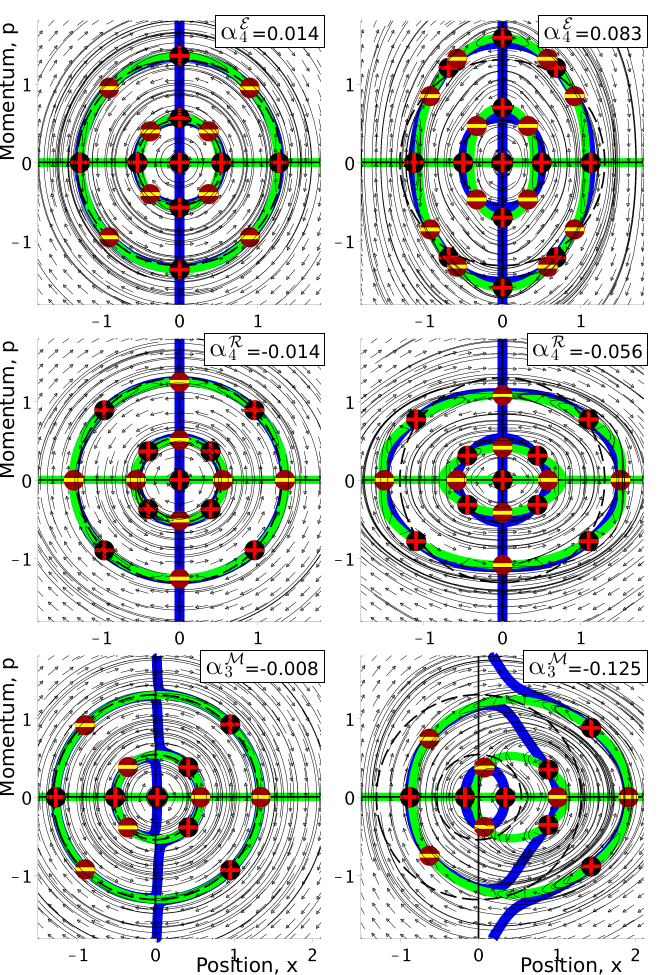}}
  \caption{\CO{(Color Online)} \emphCaption{Wigner current patterns of the first excited state (left
      frame) and second excited state (right frame)} for the three classes of weakly-anharmonic
    potentials: top row, (hard) Eckart, middle row, (soft) Rosen-Morse, bottom row, (odd)
    Morse. Note that the current patterns of odd potentials feature shapes of the hard case~(for
    $x<0$) and the soft case~(for $x>0$). The anharmonicity~$\alpha_\nu^{\cal A}$
    in~Eq.~(\ref{eq:V_trunc}) increases from left to right, with the respective values quoted in
    each panel, corresponding to values of $D=$~(12, 2)~(top row), (12, 3)~(middle row), and (2048,
    8)~(bottom row).  Symbols, linestyles and their coloring have been adopted from
    fig.~\ref{fig:current_harm}~(dashed black lines show zeros of~$W^\odot$). The stagnation points
    at the thick zero-lines' crossings are labelled as in
    fig.~\ref{fig:stagnation_points_omega}. All fieldlines form closed
    loops,~$\bm \nabla \cdot \bm J=0$, since for eigenstates $\partial_t W
    =0$. \label{fig:zeros_anharm_1}}
\end{figure*}

The \emph{degeneracy} of Eq.~(\ref{eq:WF_harm}) leads to formation of \emph{lines} of
stagnation~\cite{Ole_PRL13} in the harmonic case. When anharmonicities are `turned on'
($\alpha_\nu>0$), the zero-lines of the respective current components $J_x$ and $J_p$ get
shifted by different amounts. This results in the formation of stagnation \emph{points} in
phase-space where the two components' zero-lines cross; compare
fig.~\ref{fig:current_harm} with~\ref{fig:prediction_zeros_anharm}
and~\ref{fig:zeros_anharm_1} where thick green lines show where $J_x=0$ and thick blue
lines where $J_p=0$.

An intuitive understanding of the ensuing Wigner current patterns is discussed next.

\subsection{Existence of distinct Stagnation Points of Wigner
  Current \label{subsec_WF_points}}

The anharmonicity of the potential deforms the zero-contours of~$W$ shifting
the zero-lines of the $J_x$-component
($J_x=\frac{p}{M}W$). The $J_p$-zero-lines
get shifted differently due to the additional presence of the quantum corrections terms in
Eq.~(\ref{eq:CurrentComponents}): anharmonic quantum-mechanical systems form discrete
current stagnation points in phase-space whenever $\hbar>0$.

In short, weakly-anharmonic systems are fundamentally
non-classical~\cite{Oliva_Traj1611,Kakofengitis_PRA17}. The Wigner distributions for
energy eigenfunctions of a weakly-anharmonic system converge pointwise towards those of
the harmonic oscillator, but~$\bm J$ and its fieldlines do not. In this sense there cannot
be a smooth transition from quantum to classical case in either the limit of~$\hbar
\rightarrow 0$ or vanishing non-linearity in the potential.

This is at variance with published statement such as --- \emph{``Trajectory methods [...]
  are not reliable in general, being restricted to interaction potentials which do not
  deviate too much from an harmonic potential.''}~\cite{Daligault_PRA03}, or:~\emph{``the
  first step toward a systematic and general Wigner description is to consider a system
  whose potential differs only slightly from a harmonic
  potential''}~\cite{Lee_Scully_JCP82}.

Instead, we find that very weakly anharmonic quantum systems develop quantum coherences
essentially just like more strongly anharmonic systems, only more
slowly.

\subsection{Qualitative Effects of Anharmonicities: features of eigenstates' current
  stagnation points \label{subsec:QualitativeAnharmonic}}

Heisenberg's uncertainty principle $\Delta x \cdot \Delta p \geq \hbar /2$ implies
constancy of the size of an uncertainty domain in
phase-space~\cite{Walls_Milburn_QuopBook} (note that this argument must not be taken too
far~\cite{Zurek_NAT01}).  Hard potentials squash phase-space fieldlines in position thus
elliptically expanding them in momentum, see bottom row of fig.~\ref{fig:Potential}. This
observation can be applied to the shape of $W$'s
zero-circles~(Section~\ref{subsec:hosc_J_degenerate}) as well: compare the green lines in
fig.~\ref{fig:prediction_zeros_anharm} and in the top row of
fig.~\ref{fig:zeros_anharm_1}. Soft potentials invert this scenario, expansion in~$x$
leads to an elliptical squeeze in~$p$, see middle row of
fig.~\ref{fig:zeros_anharm_1}. Odd potentials are effectively hard on the left and soft on
the right side. This leads to a growth in position spread and reduction in momentum
spread, similar to the case of soft potentials; but, additionally, phase-space features
tend to be moved to the right, towards the side where the potential is open, see bottom
row of fig.~\ref{fig:zeros_anharm_1} and our discussion on the displacement of the minimum
vortex in section~\ref{paragraph_Displ_min_vortex}.

The $x$-axis is colored green to mark the vanishing of the component~$J_x$, yielding two
stagnation points for all (blue) $J_p$-zero-circles intersecting it.  Similarly, the
$p$-axis is a blue line in the harmonic case, and, for symmetry reasons, also for even
potentials. For odd potentials these $J_p$ zeros do not lie on the $p$-axis but are
displaced to the right, see bottom row in fig.~\ref{fig:zeros_anharm_1}.

In section~\ref{paragraph_Displ_Jp_Jx} we confirm these
statements through a mathematical analysis. 

Can an alternative to the break-up of the $J_p$ zero-lines for odd potentials, see bottom
row of fig.~\ref{fig:zeros_anharm_1}, exist? 

The answer is --it cannot: to the left of the $p$-axis an odd potential is hard and therefore has to
yield the charac\-teristic pattern displayed in the top row of fig.~\ref{fig:zeros_anharm_1}, to the
right it is soft, yielding the middle row pattern. Near the $p$-axis both patterns meet but cannot
be connected due to the continuity of $J_x$ and $J_p$ as functions of $x$ and $p$. The only option,
respecting continuity, is the cut-and-reconnect pattern we see realised in
bottom row of fig.~\ref{fig:zeros_anharm_1}.

In the limit of vanishing anharmonicity, four stagnation points form on the diagonals $|x|=|p|$ per
zero-circle. These positions can be understood from the above observations. The elliptic squashing
and expansion of the zero-circles of $J_x$ and $J_p$ is weak, leading to deformation of a
zero-circle into two ellipses with small, slightly different eccentricities, common centres and
equal area which are aligned with the coordinate axes of phase-space. In the limit of vanishing
eccentricities these intersect at odd multiples of 45 degrees forming the \emph{diagonal stagnation
  points} we observe in Figs.~\ref{fig:prediction_zeros_anharm} and~\ref{fig:zeros_anharm_1}.

To summarize this qualitative discussion: 

Weakly-anharmonic even potentials have $8n+1$ stagnation points for all low lying
eigenstates~$\psi_n$: one near the origin, 4 diagonal stagnation points and 4 stagnation points:
2, where the $J_p$-zero-lines cross the $x$-axis, and 2, where the $J_x$-zero-lines cross the $p$-axis.

Weakly-anharmonic odd potentials have $6n+1$ stagnation points per eigenstate, since the
$p$-axis stagnation points are avoided by the cut-and-reconnect mechanism, mentioned
above.

For very great anharmonicities some of our results are approximations, see top right-most
panel of fig.~\ref{fig:zeros_anharm_1}.

\subsection{For eigenstates, displacement of $J_p=0$, on the $x$-axis, is less than that of $J_x=0$ \label{paragraph_Displ_Jp_Jx}}

Numerically, we see that the zero lines of $J_x$ and $J_p$ shift differently. We now confirm
analytically our qualitative discussion in section~\ref{subsec:QualitativeAnharmonic}.

For the displacement analysis of weakly anharmonic potentials we use $\bm J$ up to first order
in~$l$, in Eq.~(\ref{eq:CurrentComponents}). Because $\alpha_\nu\ll 1$, $W \approx W^\odot$ and
$J_x\approx pW^\odot$. Furthermore
\begin{align}
  \label{eq:_Jp_in_x}
  J_p&\approx -W^\odot\partial_x V_\nu +\frac{\hbar^2}{24}\partial^2_p W^\odot\partial^3_x
  V_\nu &
  \\
  & = -xW^\odot + \frac{\hbar^2 \nu!}{24 (\nu -3)!} \cdot \alpha_\nu x^{\nu-3}
  \cdot \partial^2_p W^\odot\;,&
  \label{eq:Dx_Jp_in_x}
\end{align}
\begin{align}
\text{and } & \partial_x J_p \approx\partial_x \left( -W^\odot \partial_x V_
\nu
\right) = -W^\odot -x\partial_x W^\odot\; .&
\end{align}

We determine the displacement~$\delta x_{J_p}$ of the zeros of $J_p$ on the $x$-axis using the Newton
gradient method at $(x,p)=(\tilde X,0)$, here $\tilde X$ denotes the point where the Wigner
distribution is zero, i.e., where~$J_x=0$. Thus, $\delta x_{J_p} |_{(\tilde X,0)} \approx -
J_p|_{(\tilde X,0)} /\partial_x J_p|_{(\tilde X,0)} $, which yields
\begin{equation}
  \Big. \delta x_{J_p} \Big|_{(\tilde X,0)} \! \! \approx 
\left.
  \frac{\hbar^2 \nu!}{24 (\nu -3)!} 
  \cdot \alpha_\nu  x^{\nu-4} 
        \cdot 
\frac{\partial^2_p W^\odot}{ \partial_{x} W^\odot} 
      \right|_{(\tilde X,0)}
   \! \!  .\label{eq:_Delta_Jp_X0}
\end{equation}

Now, the Mexican hat profiles of the harmonic oscillator's Wigner distributions (see insets
fig.~\ref{fig:current_harm}) imply that $\partial^2_p W_{n,n}^\odot / \partial_x W_{n,n}^\odot
|_{(\tilde X,0)}$ is positive for $\tilde X > 0$ and negative for 
$\tilde X < 0$.

For example, if the potential is stiff and symmetric ($\alpha_4 >0$) we know that the contours of
$W$ are squeezed inward on the $x$-axis, due to the uncertainty principle. In other words the
magnitude of the zeros of the Wigner distributions on the $x$-axis obey $|\tilde X_W|< |\tilde
X_{W^\odot}|$.  In this case $ \Big. \delta x_{J_p} \Big|_{(\tilde X,0)} > 0$, which counteracts the
inward movement of the zeros of $J_x$; the zero line of~$J_p$ is less deformed than that of~$J_x$.

The same logic can be applied to soft weakly-anharmonic potentials. 
Although for odd potentials a
higher order in~$l$ of Eq.~(\ref{eq:CurrentComponents}) has to be used, 
this discussion confirms
that an odd potential's behaviour constitutes a hybrid of stiff and 
soft potentials' behaviour, see
subsection~\ref{subsec:QualitativeAnharmonic} and fig.~\ref{fig:zeros_anharm_1}.

Additionally, the discussion above shows that quantum dynamics in phase space, in the case of
vanishing Planck constant $\hbar$ or vanishing anharmonicity, does not pointwise converge to
classical dynamics.

\section{Wigner current patterns for Two-State Superpositions\label{sec_6_2state_dynamics}}

\begin{figure*}[ht!]
  \centering
  \includegraphics[width=1\textwidth,angle=0]{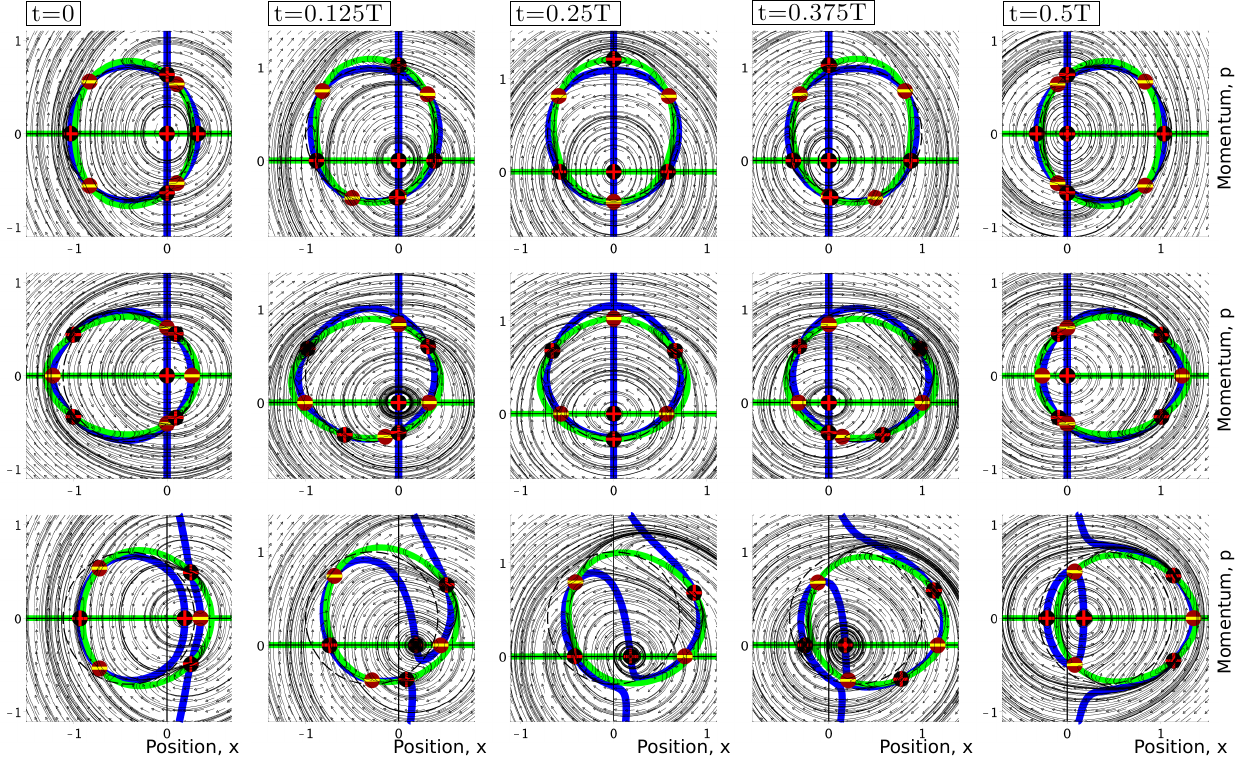}
  \caption{\CO{(Color Online)} \emphCaption{Wigner current patterns~$\bm
      J$~(\ref{eq:CurrentComponents}) for the superposition
      state~$\Psi_{0,1}(t;\frac{\pi}{3})$ for times $t/T^\mathcal{A}_{0,1} = 0$, $1/8$,
      $1/4$, $3/8$, $1/2$. The same symbols as in fig.~\ref{fig:zeros_anharm_1} are
      used~(dashed black lines show zeros of~$W^\odot$). The potentials are (see
      Table~\ref{tab:SIPs}) Eckart potential, with $\alpha_4^{\cal E}=0.042$~($D=4$) (top
      row), Rosen-Morse potential, with~$\alpha_4^{\cal R}=-0.042$~($D=4$,~$N=8$), and
      Morse potential, with~$\alpha_3^{\cal M}=-0.088$~($D=16$,~$N=32$). Note that the
      Morse potential is odd, i.e. it is hard on the left ($x<0$) and soft on the right
      side ($x>0$); accordingly, the current patterns on the left resemble those of the
      Eckart potential depicted (top row) and those on the right resemble those for the
      Rosen-Morse potential (middle row).
\label{fig:EckartImpPlots}}}
\end{figure*}

\begin{figure*}[ht!]
  \centering
  \includegraphics[width=0.96\textwidth]{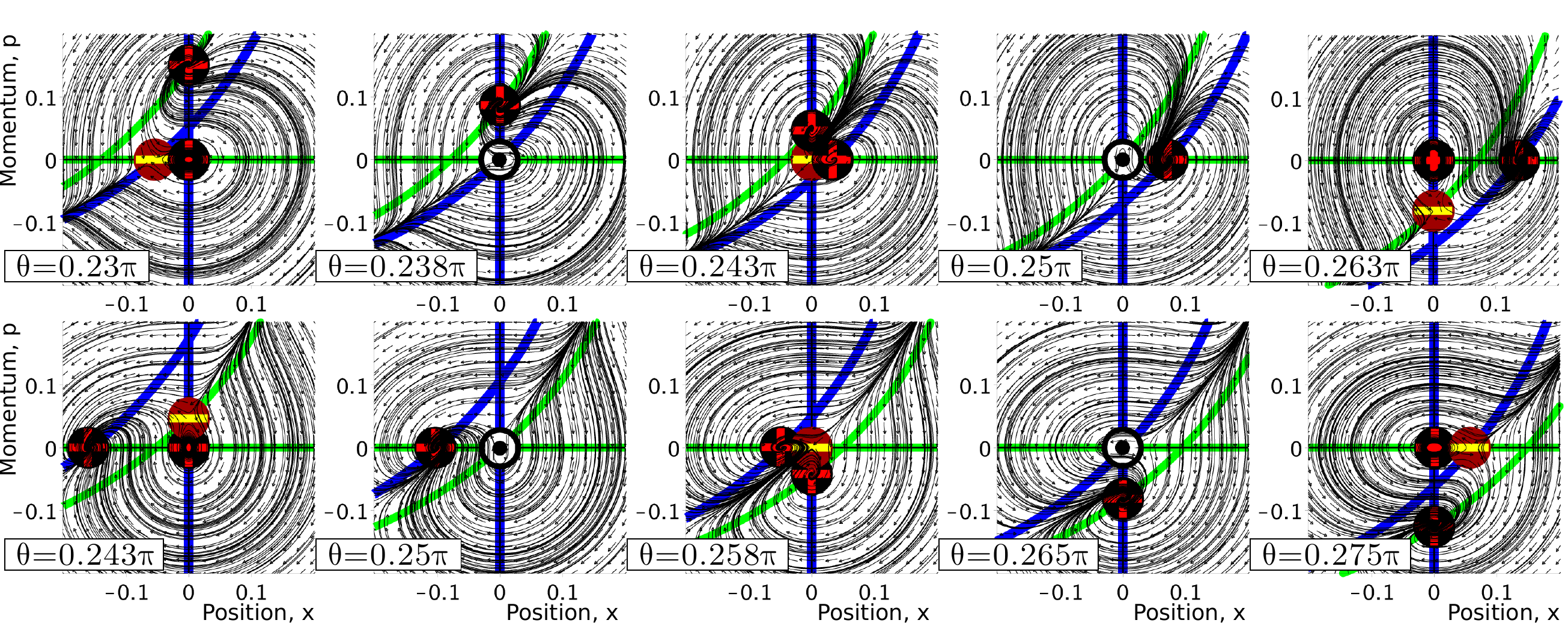}
  \caption{\CO{(Color Online)} \emphCaption{Demonstration of conservation of current
      winding number for fixed time~$t=T/8$ parameterized using a varying weighting
      angle~$\theta$ of superposition state~$\Psi_{0,1}(x,\frac{T}{8};\theta)$}~(dashed
    black lines show zeros of~$W^\odot$). We observe movements of stagnation points
    leading to their merger and splitting. The top row refers to the hard Eckart and
    bottom row to the soft Rosen-Morse potential with identical parameters as
    in~fig.~\ref{fig:EckartImpPlots}. Note the non-Liouvillian nature of the
    current~\cite{Kakofengitis_PRA17} featuring regions of pronounced expansion and
    compression.\label{fig:RosenMorse_ZoomIn}}
\end{figure*}

We now consider superpositions of energy eigenstates. Note that the associated fieldline patterns
presented in Figs.~\ref{fig:EckartImpPlots},~\ref{fig:RosenMorse_ZoomIn}
and~\ref{fig:zeros_2nd_Excited_state}, are integrated lines of~$\bm J$ at one moment in time
only. They therefore do not represent the time-evolution of $\bm J$, but an illustrative, albeit
somewhat unphysical, momentary snapshot.

\subsection{Displacement of the minimum vortex\label{paragraph_Displ_min_vortex}}

Similarly to Eq.~(\ref{eq:_Delta_Jp_X0}), with the Newton gradient method we we determine the
$x$-shift of the zero of $J_p$ at the origin $\delta x_{J_p} |_{(0,0)} \approx - J_p|_{(0,0)} /
\partial_x J_p|_{(0,0)}$, and find that the minimum vortex' shift is
\begin{eqnarray}
  \Big. \delta x_{J_p} \Big|_{(0,0)} \! \! \! \! & \approx  \left.
     \frac{\hbar^2 \nu(\nu-1)(\nu-2)}{24} 
          \cdot \alpha_\nu  x^{\nu-3} 
      \cdot \frac{\partial^2_p W^\odot}{W^\odot}  
  \right|_{(0,0)} .&\label{eq:_Delta_Jp_00}
\end{eqnarray}

For even potentials the stagnation point of $\bm J$ near its minimum does 
not shift at all, because 
$\partial^{(2l+1)}_x
V|_{(0)}=0$. This result conforms with our
expectation~(Section~\ref{subsec:QualitativeAnharmonic}) that, for symmetry reasons, the
vortex at the origin of eigenstates of even potentials does not shift. This can be
confirmed, to all orders in~$\alpha$, using~(\ref{eq:CurrentComponents}).

The stagnation point of $\bm J$ near the minimum of the potential only shifts for odd
potentials. If the potential is anharmonic with its leading term~$\alpha_\nu$ of higher
than third order, a higher order expansion has to be performed. With a leading third order
anharmonicity ($\alpha_3 < 0$) the Mexican hat profiles of the harmonic oscillator's
Wigner distributions (see insets fig.~\ref{fig:current_harm}) imply that $\partial^2_p
W_{n,n}^\odot / W_{n,n}^\odot |_{(0,0)} < 0$.  Therefore, according to
Eq.~(\ref{eq:_Delta_Jp_00}), with~$\nu=3$, $ \delta x_{J_p} |_{(0,0)} > 0$.  This confirms
the shift to the right, in the direction of the potential's opening, as predicted in the
qualitative discussion in Section~\ref{subsec:QualitativeAnharmonic} and visible in the
bottom row of fig.~\ref{fig:zeros_anharm_1}.

\begin{figure}[t]
  \centering
  \includegraphics[width=0.35\textwidth,height=0.33\textwidth]{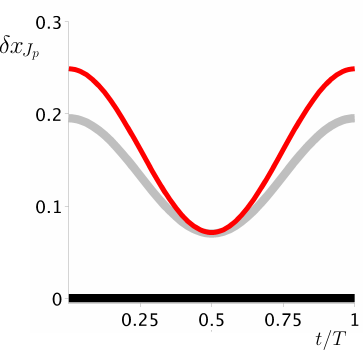}
  \caption{\CO{(Color Online)} \emphCaption{Time-dependent quantum displacement~$\delta x_{J_p}(t)$,
      Eq.~(\ref{eq:_Delta_Jp_00}) of the Morse potential minimum's stagnation point for
      superposition state $\Psi_{0,2}(x,t;\frac{\pi}{4})$. Anharmonicity value $\alpha_3^{\cal
        M}=-0.088$ ($D=16$,~$N=32$; see~Table~\ref{tab:SIPs}). Black line: position of potential's
      minimum. Red line: numerically determined displacement using $\bm J$
      of~Eq.~(\ref{eq:CurrentComponents}). Grey line: first order approximation
      Eq.~(\ref{eq:_Delta_Jp_00}), namely: {$\delta x_{J_p} \approx
        \frac{\sqrt{2}\cos\left(2t-\frac{3t}{2D}\right)+3}{4\sqrt{2D}}$}.
    \label{Fig:_min_displacement}}}
\end{figure}

For a superposition state's time-dependent displacement of the vortex near the minimum of
the potential,~$\delta x_{J_p}(t)$, Eq.~(\ref{eq:_Delta_Jp_00}) provides a reasonably good
approximation. This is depicted in fig.~\ref{Fig:_min_displacement}.

Note that in the classical case the minimum does not shift at all, compare
fig.~\ref{fig:Potential}, the shift of the minimum vortex is a pure quantum effect.


\subsection{The Ferris Wheel Effect -- alignment with $x$- and  $p$-axes \label{subsec_FerrisWheelEffect}} 

According to the discussion in Section~\ref{subsec:QualitativeAnharmonic}, four diagonal
stagnation points form per zero-circle of every eigenstate.  If we `perturb' an eigenstate
by, say, mixing in a little bit of groundstate~($\Psi_{m,0}(\theta)$ of
Eq.~(\ref{eq:superposition_state}) with $\theta \ll 1$), the zero-circle gets displaced from the
origin [see Eq.~(\ref{eq:HarmCentr})].  Yet, for small values of $\theta$ the four diagonal
stagnation points remain pinned to the zero-circle while it rotates around the origin as time
progresses. They do this such that they maintain their relative orientation with respect to the axes
of phase-space, as seen from the zero-circle's centre. In other words, while they travel through
phase-space they behave somewhat like markers on a Ferris wheel cabin, where the zero-line,~$J_x=0$,
depicts the cabin's outline, see fig.~\ref{fig:EckartImpPlots}.

\subsection{Rabi scenario: modified two-state dynamics
  \label{subsec_RabiScenario}}

To investigate a simple system in which the weighting angle~$\theta$ of the superposition
state~(\ref{eq:superposition_state}) changes considerably while the dynamics progresses, we study a
resonantly driven Rabi system. Its solution for a superposition of ground and first excited state is
\begin{equation}
  \Psi^R_{0,1} (x,t; \theta(t) ) = \Psi_{0,1}(x,t; \frac{\Omega_R}{2} t + \frac{\pi}{2})  \; ,
  \label{eq:Rabi_state}
\end{equation}
where $\Omega_R$ is the Rabi frequency~\cite{Walls_Milburn_QuopBook} and the rotating wave
approximation has been used. In accord with this approximation we assume that the
perturbation is so small that we can neglect the time-dependence of the Hamiltonian when
determining the fieldlines of~$\bm J$.

\begin{figure}[t!]
  \centering
  \includegraphics[width=0.490\textwidth]{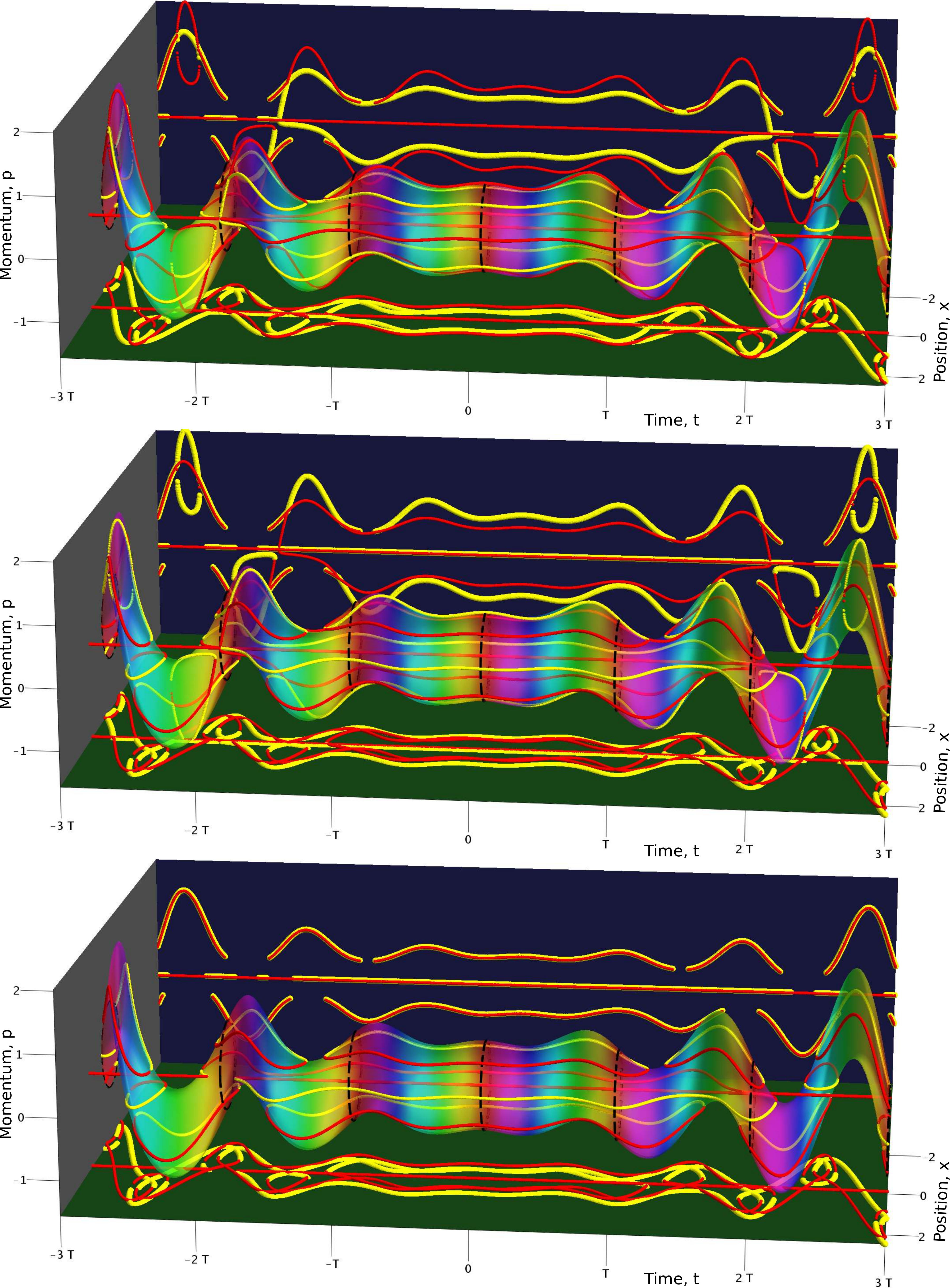}
  \caption{\CO{(Color Online)} \emphCaption{Rabi scenario for state~(\ref{eq:Rabi_state})
      with frequency ratio $\Omega_R/\Omega^\odot =1/8$.} The panels feature, from top to bottom, a soft, hard and
    odd potential,~Table~\ref{tab:SIPs}, with respective anharmonicity
    values,~$\alpha_4^{\cal E}=0.002$,~$\alpha_4^{\cal
      R}=-0.002$ and~$\alpha_3^{\cal
      M}=-0.004$. At time $t=0$ the system is in the first excited state; at other
    times the zero-circle's center~(\ref{eq:HarmCentr}) is displaced from the origin such
    that, over time, it sweeps out a helix with varying width. 
This is displayed as a helical tube whose rainbow coloring depicts the flow of time.
Every full period ($T^\odot = 1$) is
    denoted by a dashed black zero-circle. 
 Stagnation points are
    depicted by red lines when carrying charge~$\omega=+1$ and yellow if $\omega=-1$,
    compare fig.~\ref{fig:stagnation_points_omega}. The stagnation point positions are
    additionally projected along the $x$-axis onto the blue wall in the back and along the
    $p$-axis downward onto the green floor.
    Winding number conservation implies that positively and negatively charged stagnation
    points originate and annihilate together, this is seen as red and yellow lines forming
    loops which are reminiscent of the formation of the torus reported in fig.~4 of
    reference~\cite{Ole_PRL13}. As mentioned in fig.~\ref{fig:zeros_anharm_1}
     above, the bottom panel, for the odd Morse potential,
    inherits features of Rosen-Morse and Eckart potentials.
    \label{fig:SeveralRabiCycles}}
\end{figure}

The Rabi state~(\ref{eq:Rabi_state}) displays Wigner current patterns associated with the system's
(fast) intrinsic dynamics while (slowly) shifting the weighting of the superposition state: for the
ratio of these two system frequencies we choose~$\frac{\Omega_R}{\Omega^\odot} =\frac{1}{8}$ in
fig.~\ref{fig:SeveralRabiCycles}.

To monitor the effects of the slow shift of $\theta$ by itself we keep time fixed and change
$\theta$ `by hand'. The topological nature of the stagnation points conserves the current winding
number in this case as well, see fig.~\ref{fig:RosenMorse_ZoomIn}.

For the full time-dependence we choose~$\frac{\Omega_R}{\Omega^\odot} =\frac{1}{8}$ in
fig.~\ref{fig:SeveralRabiCycles}. It shows plots with zero-circles~(\ref{eq:HarmCentr}) tied to a
spiral centred on $t=0$ (since~$ \Psi^R_{0,1}(t=0)= \psi_{1}$) which expands outward as more of the
groundstate gets mixed in with increasing values of~$|t|$. We notice that the Ferris wheel-effect
tends to keep the orientation of the stagnation points on the zero circle aligned with $x$- and
$p$-axes.  With our choice of~$\frac{\Omega_R}{\Omega^\odot} = \frac{1}{8}$, around $|t|=2T$ the
mixing angle is roughly $|\theta| =\frac{\pi}{4}$. At this stage the zero-circle gets displaced by
its radius and stagnation points on the circle interact with those on $x$- and $p$-axes, see
fig.~\ref{fig:SeveralRabiCycles}, displaying repulsion, attraction, coalescence and splitting of
stagnation points-- all constrained by conservation of topological charge.

\begin{figure*}[t]
  \centering
  \includegraphics[width=0.95\textwidth]{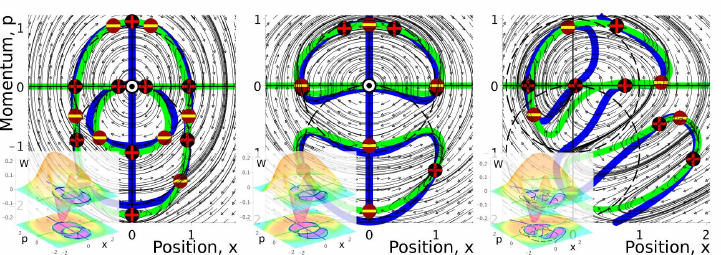}
  \caption{\CO{(Color Online)} \emphCaption{Wigner current pattern for
      state~$\Psi_{1,2}\left(x,\frac{3T}{4};\frac{\pi}{4}\right)$} for, from left to
    right, Eckart, Rosen-Morse and Morse potentials,~Table~\ref{tab:SIPs}, with respective
    anharmonicity values~$\alpha_4^{\cal E}=0.042$~($D=4$), $\alpha_4^{\cal
      R}=0.033$~($D=5$,~$N=10$), and $\alpha_3^{\cal M}=0.125$~($D=8$,~$N=16$). The same
    symbols as in~fig.~\ref{fig:zeros_anharm_1} are used~(dashed black lines show zeros
    of~$W^\odot$). Zeros of the Wigner distribution (thick green lines) intersect with
    zeros of the momentum component of Wigner current (thick blue line) yielding fairly
    intricate arrangements of stagnation points of Wigner current. Similarly to
    Figs.~\ref{fig:zeros_anharm_1} and~\ref{fig:SeveralRabiCycles}, the Morse potential's
    case inherits features of a stiff potential for $x<0$ and of a soft potential for
    $x>0$. \label{fig:zeros_2nd_Excited_state}}
\end{figure*}

\subsection{Other Superpositions \label{subsec_high_order_super}}

Other superposition states, such as $\Psi_{1,2}$, can show symmetric flower petal
arrangements, see insets in fig.~\ref{fig:zeros_2nd_Excited_state}, which have recently
been observed experimentally~\cite{Hofheinz_NAT09}. fig.~\ref{fig:zeros_2nd_Excited_state}
shows how the three different types of weakly-anharmonic potentials give rise to current
patterns which generalise our previous discussions in
Sections~\ref{subsec:QualitativeAnharmonic} and~\ref{subsec_FerrisWheelEffect}.

\section{Motivation and conclusion\label{sec_7_conclusions}}

Our investigations of Wigner current~$\bm J$ and its fieldlines shows that they give us
insights into quantum phase-space dynamics:

$\bm J$-fieldlines provide visualisation at-a-glance, in this sense, their collection across phase-space
are quantum analogs of classical phase-space trajectories.

$\bm J$ and collections of its fieldlines reveal subtle patterns in phase-space dynamics,
such as contracting and expanding regions of phase-space, current stagnation points,
loops, separatrices and saddles; similar to classical phase
portraits~\cite{Cvitanovic_Chaos_book_12,Nolte_PT10}.

In contrast to classical phase space current, $\bm J$ is non-Liouvillian~\cite{Kakofengitis_PRA17},
compare e.g. fig.~\ref{fig:RosenMorse_ZoomIn}.

$\bm J$ can be characterised by its stagnation points' distribution and
Poincar\'e-Hopf-indices.

$\bm J$-fieldlines follow neither energy-contours nor Wigner
distribution-contours~\cite{Oliva_Traj1611}.

They allow us to check concepts such as Wigner `trajectories' and dismiss such
concepts~\cite{Oliva_Traj1611,Kakofengitis_PRA17}.
 
Phase-space quantum mechanics is useful for approximate numerical modelling of quantum
dynamics using semi-classical approximations, particularly in theore\-tical quantum
chemistry, for a good and brief recent overview see~\cite{Koda_JCP15} and references
therein.  

Wigner current and its fieldlines allow us to benchmark approximate propagation
schemes~\cite{Donoso_PRL01,Donoso_JCP03,Cabrera_PRA15} against the full theory.

We expect that investigations of Wigner current will lead to new insights into the nature
of chaotic systems~\cite{Berry_PRS87} and quantum-classical
correspondences~\cite{Jaffe_Brumer_PRL85}.

We have shown that in the case of vanishing Planck constant $\hbar$ or vanishing anharmonicity, $\bm
J$ does not pointwise converge to classical dynamics, see Section~\ref{subsec_WF_points}.

We expect Wigner current~$\bm J$ and {collections of its fieldlines} to become a widely used
tool for the study of quantum dynamics -- similar to classical phase portraits.

\section*{Acknowledgements} 

We thank Stefan Buhmann and Alan McCall for their comments on this manuscript and we are
indebted to Georg Ritter for his careful reading, probing questions, and many
comments. O.~S. thanks Paul Brumer for many stimulating discussions.

\end{document}